\documentclass[conference]{IEEEtran}
\IEEEoverridecommandlockouts
\usepackage[numbers,sort&compress]{natbib}
\usepackage{pifont}
\usepackage{amsmath,amssymb,amsfonts}
\usepackage{algorithmic}
\usepackage{graphicx}
\usepackage{textcomp}
\usepackage{array}
\usepackage{algorithm}
\usepackage{multirow}
\usepackage{colortbl} % Required for row/cell coloring
\usepackage[table]{xcolor} % Required for defining/using colors, 'table' option recommended
\usepackage{booktabs} % Recommended for cmidrule, optional
\usepackage{textcomp}
\usepackage{geometry}
\usepackage{hyperref}
\definecolor{lightgray}{rgb}{0.9, 0.9, 0.9} % Light Gray

\def\BibTeX{{\rm B\kern-.05em{\sc i\kern-.025em b}\kern-.08em
    T\kern-.1667em\lower.7ex\hbox{E}\kern-.125emX}}
\begin{document}

\title{When Pipelined In-Memory Accelerators Meet Spiking Direct Feedback Alignment: A Co-Design for Neuromorphic Edge Computing}

%\author{\fontsize{14}{8}\selectfont Haoxiong Ren$^{1}$, Yangu He$^{2}$, Kwunhang Wong$^{2}$, Rui Bao$^{1}$,\\ Ning Lin$^{2,\dag}$, Zhongrui Wang$^{3,\dag}$, Dashan Shang$^{1,\dag}$}

 \author{%
   Haoxiong~Ren\textsuperscript{1},\;
   Yangu~He\textsuperscript{2},\;
   Kwunhang~Wong\textsuperscript{2},\;
   Rui~Bao\textsuperscript{1},\;
   Ning~Lin\textsuperscript{2,$\dagger$},\;
   Zhongrui~Wang\textsuperscript{3,$\dagger$},\;
   Dashan~Shang\textsuperscript{1,$\dagger$}\\[0.6ex]
     \textsuperscript{1}State Key Laboratory of Fabrication Technologies for Integrated Circuits,\\%
     Institute of Microelectronics, Chinese Academy of Sciences, Beijing, China\\
     \textsuperscript{2}Department of Electrical and Electronic Engineering,%
     The University of Hong Kong, Hong Kong, China\\
     \textsuperscript{3}School of Microelectronics,%
     Southern University of Science and Technology, Shenzhen, China\\[0.6ex]
     $\dagger$\,Corresponding authors:\;
     linning@hku.hk,\;
     wangzr@sustech.edu.cn,\;
     shangdashan@ime.ac.cn
}
\maketitle

\begin{abstract}
Spiking Neural Networks (SNNs) are increasingly favored for deployment on resource-constrained edge devices due to their energy-efficient and event-driven processing capabilities. However, training SNNs remains challenging because of the computational intensity of traditional backpropagation algorithms adapted for spike-based systems. In this paper, we propose a novel software-hardware co-design that introduces a hardware-friendly training algorithm, Spiking Direct Feedback Alignment (SDFA) and implement it on a Resistive Random Access Memory (RRAM)-based In-Memory Computing (IMC) architecture, referred to as PipeSDFA, to accelerate SNN training. Software-wise, the computational complexity of SNN training is reduced by the SDFA through the elimination of sequential error propagation. Hardware-wise, a three-level pipelined dataflow is designed based on IMC architecture to parallelize the training process. Experimental results demonstrate that the PipeSDFA training accelerator incurs less than 2\% accuracy loss on five datasets compared to baselines, while achieving 1.1$\times$$\sim$10.5$\times$ and 1.37$\times$$\sim$2.1$\times$ reductions in training time and energy consumption, respectively compared to PipeLayer\footnote{This code is available at: \href{https://github.com/HershelYen/Spiking-Direct-Feedback-Alignment}{Spiking-Direct-Feedback-Alignment}}.
\end{abstract}


% Spiking Neural Networks (SNNs) have emerged as an energy-efficient computing paradigm, particularly well-suited for low-power edge applications such as autonomous driving \cite{peiArtificial2019} and medical data processing \cite{tianNew2021, doborjehPersonalised2021}. Unlike traditional artificial neural networks (ANNs), SNNs employ an event-driven approach, activating computations only when input spikes are received. This biologically inspired mechanism enables substantial reductions in both energy consumption and computational overhead, making SNNs particularly attractive for real-time decision-making systems \cite{peiArtificial2019}. 

\section{Introduction}
\begin{figure*}[t]
    \centering
    \includegraphics[width=\linewidth]{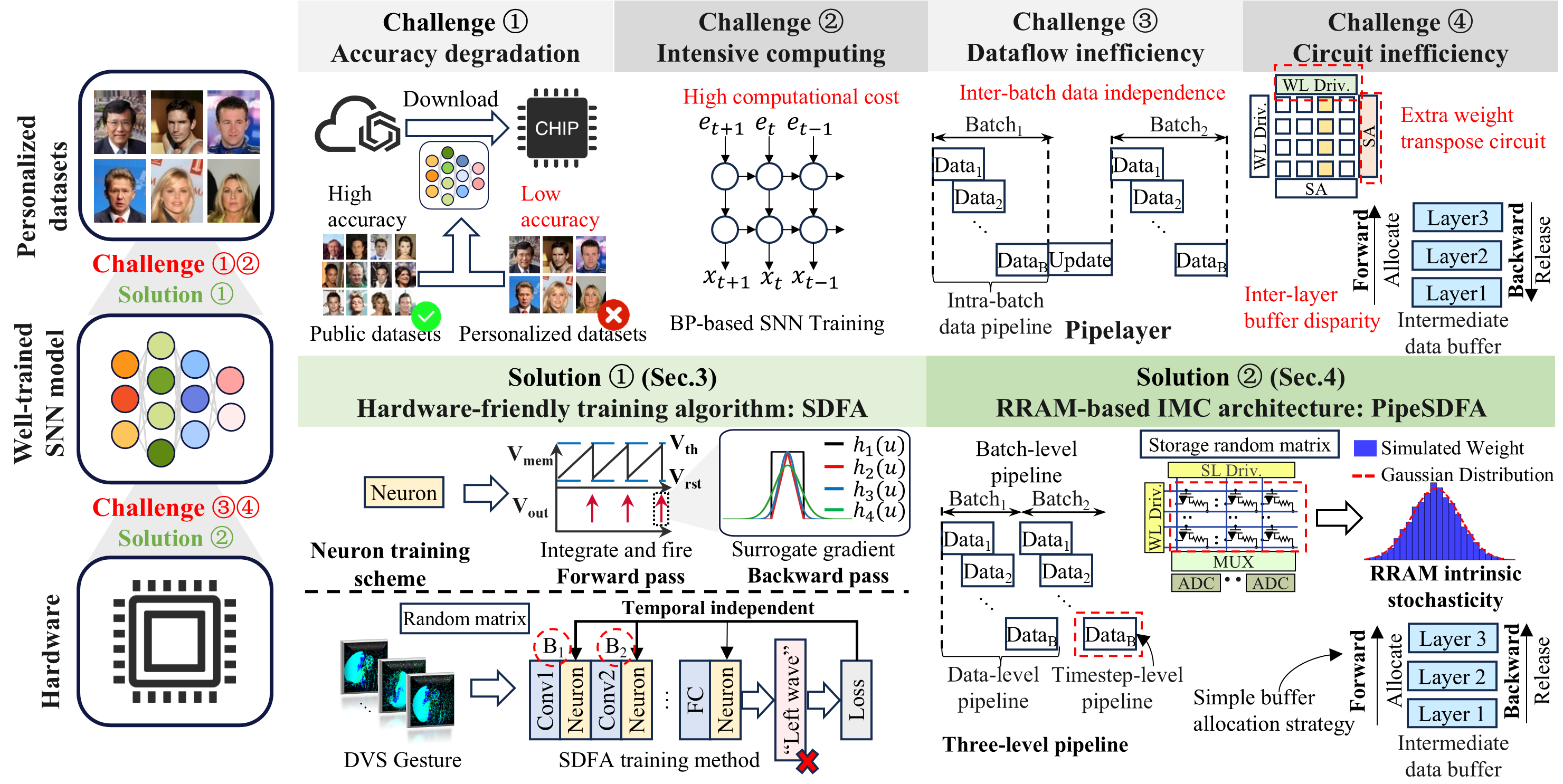}
    \vspace{-16pt}
    \caption{Challenges faced by training SNNs on edge device: \ding{172} Accuracy degradation when pre-trained model is applied to personalized datasets. \ding{173} Intensive computing arising from BP. \ding{174} Dataflow inefficiency due to sequential error propagation. \ding{175} Circuit inefficiency due to extra circuit and memory management. And solutions we provide: \ding{172} From software perspective, SDFA designed to solve the weight transpose and backward locking problem and alleviate intensive computing. \ding{173} From hardware perspective, PipeSDFA based on RRAM IMC architecture and refined design of dataflow.} 
    \label{fig:introduction}
    \vspace{-10pt}
\end{figure*}

Spiking Neural Networks (SNNs) have emerged as an energy-efficient computing paradigm, ideal for low-power edge applications like autonomous driving \cite{peiArtificial2019} and medical data processing \cite{tianNew2021, doborjehPersonalised2021}. Unlike traditional artificial neural networks (ANNs), SNNs activate computations only when input spikes occur, reducing energy consumption and computational overhead. This makes SNNs particularly suitable for real-time decision-making systems \cite{peiArtificial2019}. 
However, pre-trained SNN models often struggle to perform inference tasks on personalized datasets, as real-world user scenarios tend to be biased. While on-device training presents a potential solution, implementing efficient SNN training for personalized datasets on resource-constrained edge devices remains a significant challenge, as illustrated in Fig. \ref{fig:introduction}.

\textbf{Challenge \ding{172}: Accuracy degradation.} Although pre-trained SNNs exhibit strong generalization on standardized benchmarks, their accuracy often declines when deployed on real-world personalized datasets. This decline stems from environmental variations that deviate from pre-training data distributions. Furthermore, common deployment techniques like quantization and pruning can further reduce accuracy \cite{LNPU}. These challenges underscore the need for on-device training to adapt SNNs to diverse and dynamic contexts.

\textbf{Challenge \ding{173}: Intensive computing.} Surrogate gradient methods based on backpropagation (BP) \cite{bp} are currently the most viable approaches for training SNNs. However, the high computational demands pose challenges for deployment on resource-constrained edge devices. Additionally, the temporal dynamics of spikes introduce sequential dependencies, further exacerbating these issues.

\textbf{Challenge \ding{174}: Dataflow inefficiency.} Most current research prioritizes accelerating inference over training \cite{ISAAC,linResistive2025,inference}. While efforts like PipeLayer have improved training efficiency through pipelining and parallelism \cite{pipelayer,yuRRAM2021}, these methods struggle when applied to SNNs. The challenges stem from high memory demands and layer-by-layer dependencies in error propagation. In batch processing, the backward locking problem forces subsequent batches to wait for weight updates from prior ones \cite{locking}, a limitation that worsens with deeper networks, smaller batch sizes \cite{pipelayer} or longer timesteps \cite{hanExtension2020}. For example, PipeLayer's performance drops nearly tenfold when the batch size is reduced from 64 to 4 (see Fig. 9). In unified training-inference scenarios (i.e., batch size = 1), pipelining offers almost no performance improvement.

%\caption{Challenges faced by Training SNNs on edge device: \ding{721}}

\textbf{Challenge \ding{175}: Circuit inefficiency.} Traditional training accelerators face two key challenges. First, they require dedicated weight transpose circuits, which incur significant area and energy overheads \cite{transpose1, transpose2, transpose3}. For instance, the design in Ref. \cite{transpose1} doubles RRAM array power consumption by adding extra drivers and sense amplifiers for transpose-read operations. Second, inconsistent inter-layer buffer usage complicates memory management during training \cite{pipelayer}. Earlier network layers exhibit the longest buffer residency (allocated first and released last), creating temporal mismatches that demand complex buffering schemes such as layer-wise circular buffers to handle data lifetimes efficiently \cite{pipelayer}.

\textbf{Potential solution.} Direct Feedback Alignment (DFA) \cite{dfa} has emerged as a promising candidate, employing fixed random feedback weights to propagate errors directly from output to intermediate layers \cite{dfa}. This approach inherently avoids backward locking and weight transport issues \cite{dfa, dfa_deeper, LNPU}, thus offering potential hardware simplification. While preliminary efforts have adapted DFA to SNNs \cite{snn_dfa1, snn_dfa2, snn_dfa3, snn_dfa4}, existing methods either introduce excessive complexity or are ineffective on complex datasets. What's more, no edge neuromorphic accelerators utilizing DFA for SNN training have yet been demonstrated in practice. This gap underscores the urgent need for hardware-efficient DFA variants optimized for SNNs, as well as dedicated accelerator architectures tailored for edge deployment.

\textbf{Our Solution.} To overcome these challenges, we propose a co-designed software-hardware framework for efficient SNN training. On the software side, we develop Spiking Direct Feedback Alignment (SDFA), a novel adaptation of DFA specifically optimized for SNNs (Solution \ding{172} in Fig. \ref{fig:introduction}). On the hardware side, we design PipeSDFA, an RRAM-based IMC architecture (Solution \ding{173} in Fig. \ref{fig:introduction}) that accelerates training. The innovations are as follows:

\begin{itemize}
    \item \textbf{Software-Hardware Co-Design.} We identify four critical challenges that limit efficient SNN training on edge devices. To address these limitations, we propose a novel software-hardware co-design framework optimized for training SNNs on resource-constrained platforms.
    \item \textbf{Hardware-Friendly Algorithm.} We leverage the surrogate gradient method to overcome the non-differentiability of spiking neuron actions. Additionally, we eliminate the temporal dimension typically considered in SNN error signals. This modification significantly reduces both the time complexity (computational cost) and space complexity (memory footprint) associated with error calculation and storage.
    \item \textbf{Novel Hardware Architecture.} We design the PipeSDFA architecture, featuring a Timestep-Data-Batch three-level pipeline. The RRAM array efficiently stores the fixed random feedback matrices and exploits RRAM intrinsic stochasticity to achieve an effective Gaussian distribution initialization for these weights. Furthermore, we introduce a simple, balanced buffer scheme that capitalizes on the properties of SDFA.
    \item \textbf{Comprehensive Validation.} We validate the effectiveness of the proposed SDFA and PipeSDFA across various spiking models using five diverse datasets. Our approach demonstrates significant computational and storage efficiency improvements, achieving a 1.1$\times$$\sim$10.5$\times$ speedup and up to 1.37$\times$$\sim$2.1$\times$ better energy efficiency compared to PipeLayer, while maintaining accuracy loss within 2\% of the baseline.
\end{itemize}

\section{preliminaries and motivation}\label{sec:preliminaries}

\subsection{Spiking Neurons and Training Scheme}
 In SNNs, the spiking neurons can be defined by three processes, that is, charging, firing and resetting. A widely used spiking neuron is the Leaky Integrate-and-Fire (LIF) neuron \cite{lif1}. In addition to integrating inputs, the LIF neuron's membrane potential gradually decays over time in the absence of input spikes. This behavior is captured by introducing a leak term in the charging dynamics,
 \begin{equation}
     C \frac{dV_{mem}(t)}{dt}=I_{ext}-I_L,
     \label{charging}
\vspace{-3pt}
 \end{equation}
where $C$ is the membrane capacitance, $V_{mem}(t)$ is the membrane potential at time $t$, $I_{ext}$ is the external input current, and $I_L$ is the leak current.

The fire and reset process is described as,
\begin{equation}
    V_{out}(t) = \begin{cases}
                1 & V_{mem}(t) \geq V_{th} \\
                0 & otherwise
                \end{cases} ,\label{firing}
\vspace{-3pt}
\end{equation}
which represents once $V_{mem}$ exceeds a certain threshold $V_{th}$, the neuron fires and $V_{mem}$ is reset to zero.

\begin{figure}[t]
  \centering
  \includegraphics[width=\linewidth]{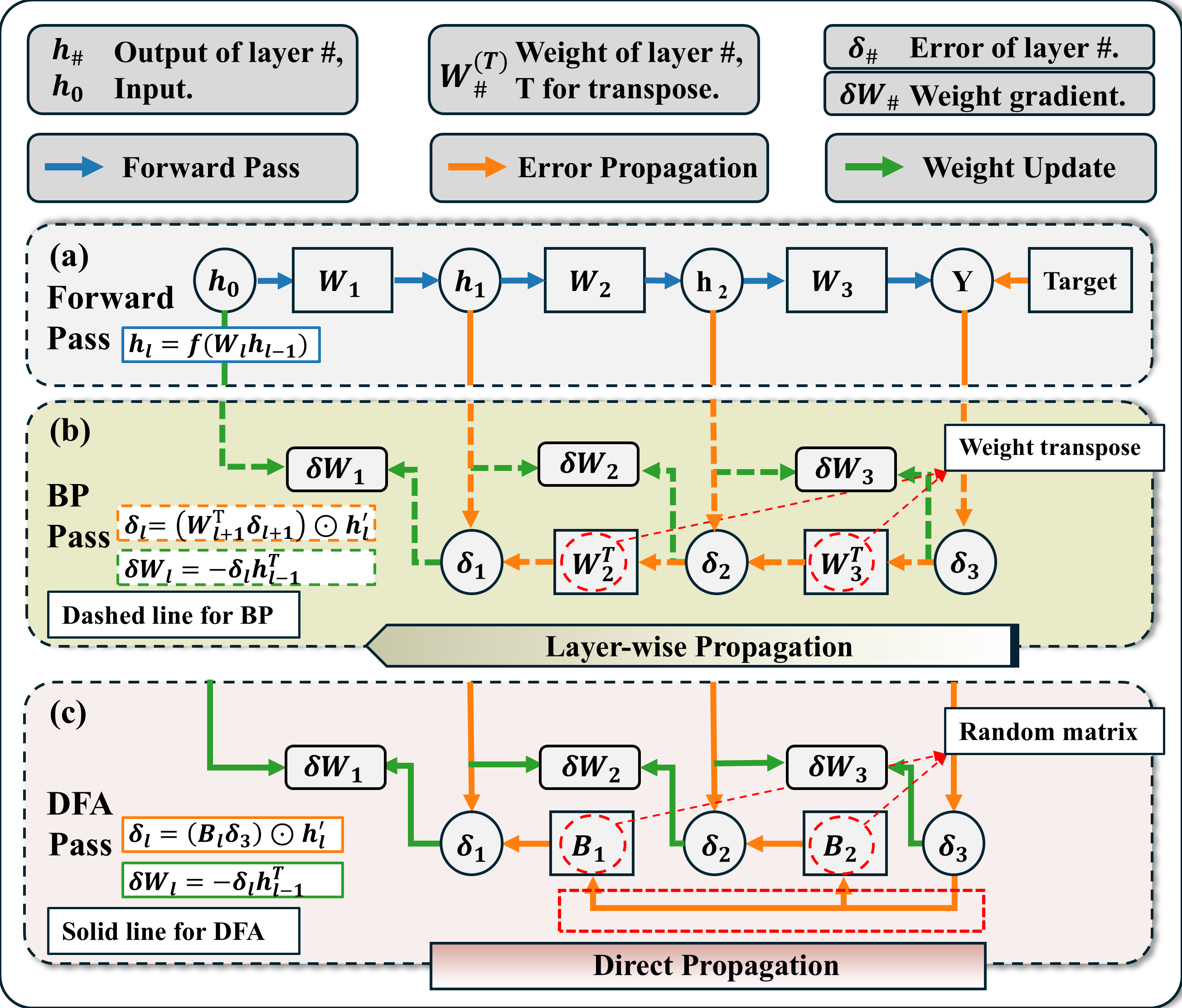}\\
  \vspace{-10pt}
  \caption{Data dependency of BP vs. DFA in ANN training process.}
  \label{ann_bp_dfa}
  \vspace{-10pt}
\end{figure}

Currently, the primary methods for SNN training are Spike-Timing-Dependent Plasticity (STDP) \cite{stdp1,stdp2}, ANN-to-SNN conversion \cite{ann2snn}, and surrogate gradient approaches \cite{surrogate2, surrogate3}. STDP is not scalable for complex tasks due to its limited handling of diverse datasets. ANN-to-SNN conversion simplifies training but often leads to performance degradation and unresolved parameter update challenges in dynamic environments \cite{ann2snn}. Surrogate gradient methods, based on BP, are preferred for their superior accuracy. However, the temporal dynamics of input spikes introduce sequential dependencies, which involve unrolling network dynamics across multiple timesteps and aggregating weight gradients at each step, resulting in significant memory overhead. This is detrimental to resource-constrained edge devices.

This work focuses on addressing the challenges posed by these training methods.

\subsection{BP vs. DFA}
\textbf{BP.} As illustrated in Fig. \ref{ann_bp_dfa}(a)-(b), the BP training process consists of three key steps: forward pass, error backpropagation, and weight update. Error signals start at the output layer and propagate backward layer-by-layer. The gradient of the loss function, computed using the chain rule, is propagated through layers, creating strict inter-layer dependencies as shown in Fig. \ref{ann_bp_dfa}(b). The error of layer $l$ is mathematically given by
\begin{equation}
    \delta_l = \left( W_l^T \delta_{l+1} \right) \odot f^{\prime}(h_{l}) \quad l \in \{1, 2, \dots, L-1\},
    \label{bp}
\vspace{-3pt}
\end{equation}
where $W^T$ represents the transposed weight, $\odot$ denotes element-wise multiplication, and $f'()$ is the derivative of the nonlinearity function used in the forward pass. This process presents two significant challenges:

\underline{\ding{172} Transpose of weight matrices.} BP requires transposed weight matrices at each layer to compute gradients, as shown by the red dashed circle in Fig. \ref{ann_bp_dfa} (b). Storing these transposed matrices increases memory footprint during training and can disrupt data locality and cache efficiency, which are critical for hardware optimization \cite{hanExtension2020, pipelayer}.

\underline{\ding{173} Backward locking.} In BP, error signals must propagate sequentially through layers, adjusting weights at each layer based on the error gradient. In batch processing, all data within a batch share the same weights, and the next batch must wait for the weight update, leading to the backward locking problem \cite{pipelayer}. This issue worsens when scaling to deeper networks or mini-batch processing.

% \underline{\ding{172} Transpose of weight matrices.} BP requires the use of transposed weight matrices at each layer to compute the gradient, as shown by the red dashed circle in Fig. \ref{ann_bp_dfa}. Storing these transposed weight matrices significantly increase the memory footprint during training, and can disrupt data locality and cache efficiency, which are critical factors for hardware optimization \cite{hanExtension2020, pipelayer}.

% \underline{\ding{173} Backward locking.} In BP, error signals must be propagated sequentially through layers, adjusting weights at each layer based on the error gradient. In batch processing, all data within a batch share the same weights, and the next batch must wait for the weight update, leading to a backward locking problem \cite{pipelayer}. This issue becomes even worse when scaling to deeper networks or mini-batch processing.

\textbf{DFA}, a bio-plausible training method, addresses these challenges by eliminating the need for error propagation through layers and avoiding the transposition of weight matrices \cite{dfa}, as illustrated in Fig. \ref{ann_bp_dfa}(c). DFA maintains the same forward pass as BP, but computes the error for each layer directly using the global error $\delta_L$ as,
\begin{equation}
    \delta_l = \left( B_l \delta_{L} \right) \odot f^{\prime}(h_{l}) \quad l \in \{1, 2, \dots, L-1\},
\vspace{-3pt}
\end{equation}
where $B_l$ is randomly initialized and remains fixed during training. This direct propagation mechanism significantly simplifies the computational process by reducing intermediate result storage requirements. This characteristic makes DFA an ideal algorithmic foundation for building efficient neural network training accelerators, particularly well-suited for in-memory computing hardware architectures.

\section{Spiking Direct Feedback Alignment}\label{sec:sdfa}

We introduce a hardware-friendly training paradigm called SDFA, which maintains competitive accuracy while significantly reducing computational complexity (Alg. \ref{alg:sdfa}).

\textbf{Initialization (lines 1$\sim$4 in Alg. \ref{alg:sdfa}).}
The algorithm begins by initializing fixed random feedback matrices $B_l \in \mathbb{R}^{n_l \times n_L}$ for each layer $l$, where $n_l$ and $n_L$ represent the dimensions of layer $l$ and the output layer, respectively. The dimensions of the feedback matrix $B_l$ for a given layer $l$ are determined by the shape of that layer's output gradient tensor. Taking a convolutional layer as an example, the gradient tensor typically includes temporal and batch dimensions, resulting in a shape such as $(T, N, C_l, H_l, W_l)$. $B_l$ is constructed to operate independently of the specific timestep and batch instance. Therefore, its dimensions are derived by omitting the $T$ and $N$ dimensions from the gradient tensor shape as:
\begin{equation}
    shape(B_l) = (C_l \times H_l \times W_l, n_L),
\vspace{-3pt}
\end{equation}
the same design principle applies to the fully connected layers as well.
These matrices are then sampled from a zero-mean Gaussian distribution:
\begin{equation}
    B_l \sim \mathcal{N}(\mu, \sigma^2), \quad \text{where} \quad \mu = 0,
\vspace{-3pt}
\end{equation}
where $\sigma$ is a standard deviation typically set to 1. The exponential decay property of the Gaussian distribution naturally bounds the weight magnitudes, ensuring network stability during training.

\textbf{Forward pass and loss calculation (lines 5$\sim$12 in Alg. \ref{alg:sdfa}).} During the forward pass, input data is processed sequentially through layers across discrete timesteps, mirroring the standard inference process. The final output $h_L$ is typically obtained by averaging the outputs over the entire time window. The difference between the network's final output and the desired target is quantified using the cross-entropy loss, which can be computed as:
\begin{equation}
    \mathcal{L} = -\frac{1}{N} \sum_{i=1}^{N} \sum_{c=1}^{C} y_{ic} \log(\hat{y}_{ic}),
    \vspace{-3pt}
\end{equation}
where $N$ denotes the total number of data points in the batch, $C$ represents the total number of distinct classes, $y_{ic}$ represents the target for data point $i$, and $\hat{y}_{ic}$ represents the model's predicted probability.

\textbf{Error propagation and weight update (lines 13$\sim$24 in Alg. \ref{alg:sdfa}).} In conventional DFA, the backward pass is structured as a fully-connected layer, irrespective of the forward pass's computational paradigm. This results in a fixed matrix multiplication methodology for error propagation \cite{hanExtension2020}. When DFA is applied to SNNs, it leads to a large memory footprint due to the connections between each output neuron and every timestep of inter-layer activity. The key innovation of SDFA lies in its error propagation mechanism, which fundamentally differs from that of DFA. SDFA decouples the size of the random feedback matrix from the number of timesteps by reducing the dimensionality of the feedback matrix, reusing $B_l\delta_L$ across timesteps. This strategy enables SDFA to retain the key advantages of DFA, such as eliminating weight transposition and addressing the backward locking problem, while significantly reducing the temporal dependency overhead associated with error propagation over time.

\section{PipeSDFA} \label{sec_arc}

\begin{figure*}[ht]
    \centering
    \includegraphics[width=\linewidth]{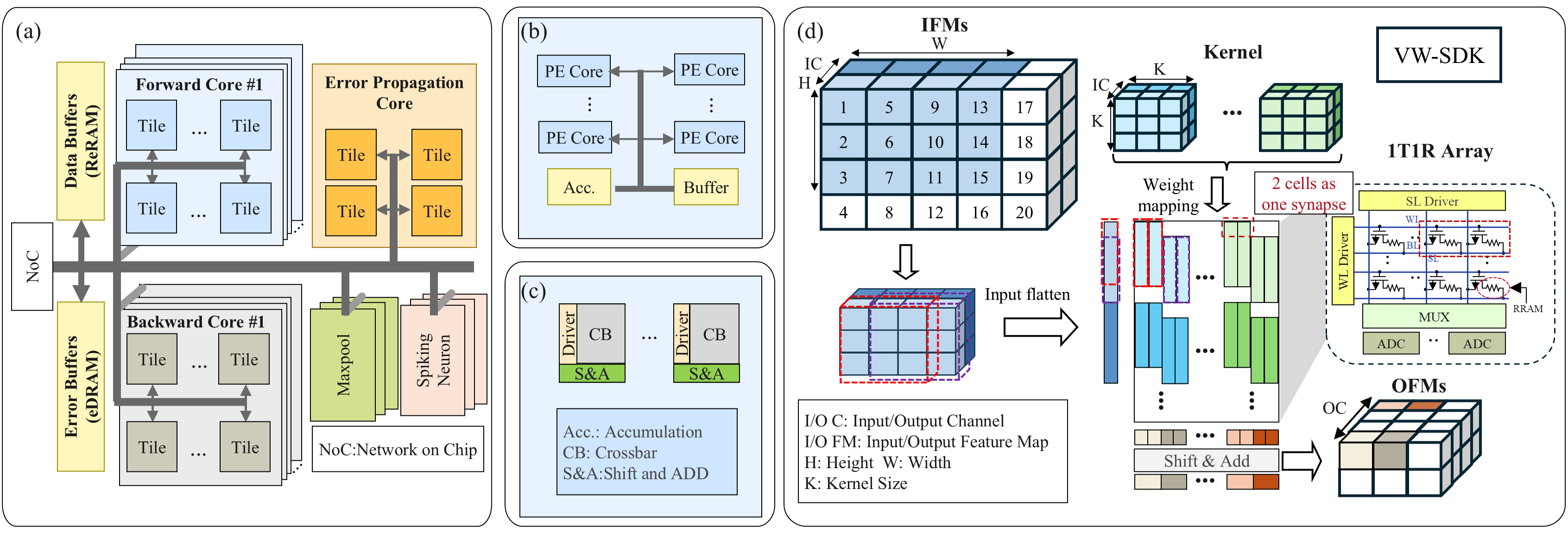}
    \vspace{-20pt}
    \caption{The hierarchy of PipeSDFA. (a) Architecture level design with NoC, buffers and computational cores. (b) Tile level design with PE Cores. (c) PE level design with crossbars. (d) VW-SDK weight mapping strategy.}
    \label{fig:arch.}
    \vspace{-6pt}
\end{figure*}

\begin{figure}[t]
    \centering
    \includegraphics[width=\linewidth]{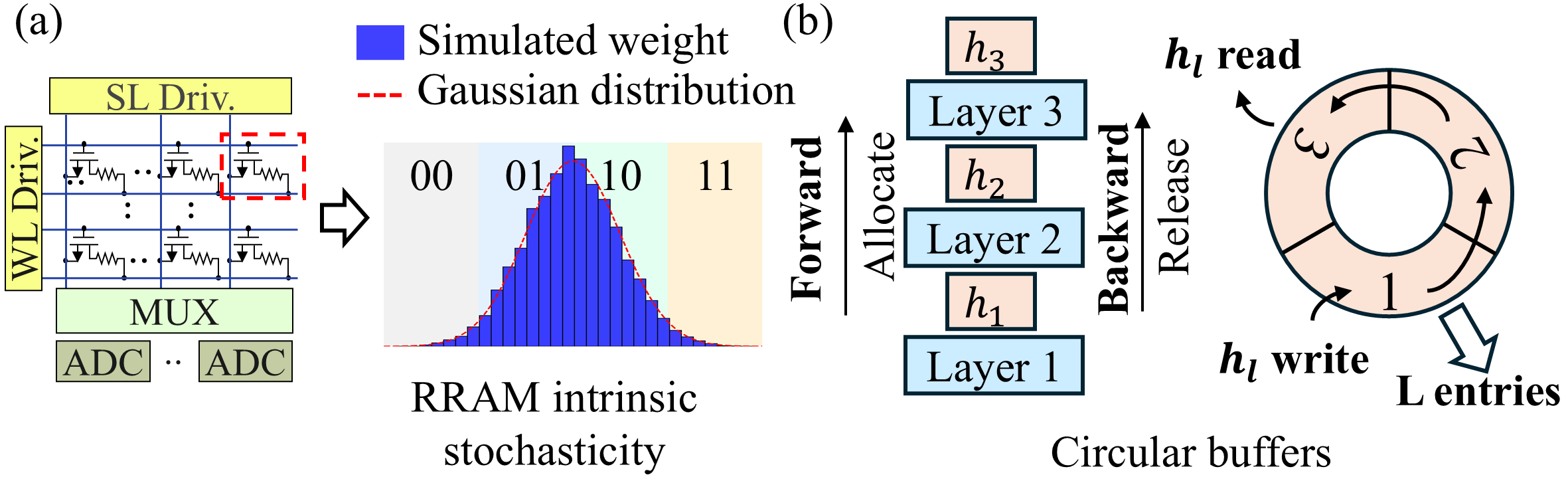}
    \vspace{-20pt}
    \caption{(a) Random matrices generation strategy using RRAM intrinsic stochasticity. (b) Intermediate data buffer design.}
    \label{fig:rram_sim}
    \vspace{-6pt}
\end{figure}

\subsection{In-memory Training Architecture}

As illustrated in Fig. \ref{fig:arch.}, we design a hardware accelerator for the proposed SDFA called PipeSDFA.

\textbf{Architecture level design.}
The Forward Core handles the primary computation of spike forward calculation through the network layers, utilizing RRAM crossbars for energy-efficient vector-matrix multiplication (VMM) operations.  Intermediate results are stored in RRAM-based Data Buffers, designed to support the temporal nature of SNN computations while minimizing data transfer overhead. The Error Propagation Core implements the novel SDFA algorithm, generating layer-wise error signals through a streamlined process that eliminates the need for weight transposition and sequential error propagation. The Backward Core completes the training loop by computing weight gradients from the error signals stored in dedicated Error Buffers. After the accumulated weight gradients are calculated, the Backward Core will collect accumulated weight gradients of every steps and perform update process.

\begin{algorithm}[t]
\small                      
\caption{Spiking Direct Feedback Alignment}
\label{alg:sdfa}
\begin{algorithmic}[1]
\REQUIRE Input data $h_0$, labels $y$, parameters $\theta = \{W_l, b_l\}$
\REQUIRE Network depth $L$, neuron function $f_l()$, timesteps $T$, shape of gradient output $grad\_shape_l$
\ENSURE Updated parameters $\theta^*$
\STATE \textbf{Feedback weight initialization:}
\FOR{each layer $l$ from $l$ to $L-1$}
    \STATE $B_l \gets init(grad\_shape_l)$
\ENDFOR
\STATE \textbf{Forward Propagation:}
\FOR{each layer $l$ from $1$ to $L-1$}
    \FOR{time step $t$ from $1$ to $T$}
        \STATE $h_l^t \gets f_l(W_l h_{l-1}^t + b_l)$
    \ENDFOR
\ENDFOR
\STATE $h_L = {\Sigma h_l^t}/T$
\STATE $\mathcal{L} \gets \text{CrossEntropy}(h_L, y)$ \COMMENT{Compute loss}

\STATE \textbf{Error Propagation:}
\STATE $\delta_L = \frac{\partial \mathcal{L}}{\partial h_{L}} \odot f'_L(W_l h_{l-1}^t + b_l)$ 
\FOR{each layer $l$ from $1$ to $L-1$}
    \FOR{time step $t$ from $1$ to $T$}
        \STATE $\delta_{l}^t \gets B_{l} \delta_L \odot f'_{l}(W_lh_{l-1}+b_l)$ \COMMENT{Direct feedback}
        \STATE $dW_l^t \gets \delta_{l}^t \cdot (h_{l-1}^t)^T + dW_l^{t-1}$ 
        \STATE $db_l^t \gets \delta_l^t + db_l^{t-1}$ 
    \ENDFOR
\ENDFOR

\STATE \textbf{Parameter Update:}
\STATE $W_l \gets W_l - \eta \cdot dW_l^T, \forall l \in \{1,...,L\}$ 
\STATE $b_l \gets b_l - \eta \cdot db_l^T, \forall l \in \{1,...,L\}$ 

\end{algorithmic}
\end{algorithm}

\textbf{Tile to PE level design.}
Within a single tile, multiple PE units are utilized collaboratively to perform computing tasks. Results from individual PEs are accumulated into an output buffer, and at the architecture level, aggregated across tiles into a central buffer. Each PE comprises multiple crossbars with associated peripheral circuits, and parameters are mapped to crossbar prior to VMM operations.

\textbf{Weight mapping.}
To address the high computational costs of SNNs and maximize mapping efficiency on IMC-based accelerators, we for the first time extend the variable-window shifted and duplicated kernel (VW-SDK) mapping approach~\cite{vwsdk} to the SNN domain. Our mapping solution explores a broader design space by supporting flexible rectangular windows (e.g., 3$\times$3 or 4$\times$4) tailored to the geometry of input feature maps and the specific resource constraints of each convolutional layer. This innovation enables the PipeSDFA architecture to achieve higher crossbar utilization and computational throughput by precisely matching layer-wise requirements to available RRAM array resources. 
The total computational cycles are calculated as,
\begin{equation}
    Total Cycles = N \times AC \times AR \times \frac{B_W}{B_C},
\end{equation}
where $N$ is the number of parallel windows, $AC$ is the array column cycles, $AR$ is the array row cycles. $B_W$ is the weight bit precision, and $B_C$ is the bit precision of a single RRAM cell. As shown in Fig. \ref{fig:arch.}(d), we implement 4-bit weights using 2 RRAM cells per synapse at the crossbar level.

\textbf{Random feedback matrices generation.} As shown in Fig. \ref{fig:rram_sim}(a), we performed 1000 Monte Carlo simulations to evaluate the statistical variations during RRAM resistance programming, and observed that the resulting distribution of resistance values closely matches a Gaussian distribution. In our mapping scheme, an ideal write resistance of 10 $k\Omega$ is targeted, and the resistance range is partitioned such that each RRAM cell encodes a 2-bit weight. Compared to conventional digital circuits, this method eliminates the need for additional random number generator modules, instead utilizing the intrinsic stochasticity of the RRAM programming process to generate truly random weights. After initial programming, the random matrices remain fixed during network training.

\textbf{Intermediate data buffer storage.}
The introduction of pipeline increases the usage of cache arrays. In Pipelayer, circular buffers of varying sizes are assigned to each layer, where buffer requirement at the $l$-th layer is $2(L-l)+1$ \cite{pipelayer}. In contrast, for our architecture, the error propagation order enables a first-allocate-first-release strategy. As illustrated in Fig. \ref{fig:rram_sim}(b), for example, the output $h_1$ from the first layer is written into the buffer and will be read three cycles later to compute the gradient for $W_2$. Immediately after this, the buffer can be safely overwritten with new data. All layers require the same buffer size $L$ resulting in a much simpler and balanced buffer allocation scheme across the network.

\subsection{Three-level Pipelined Dataflow}

% \textbf{Impact of SNN Characteristics on PipeLayer Pipeline Efficiency.}
% PipeLayer enhances networks performance through both intra-layer and inter-layer parallelism \cite{pipelayer}. Inter-layer parallelism is achieved by pipelined architecture, which significantly improves parallelism as the batch size increases. However, SNNs, which are based on time sequences, require additional computational resources and memory to handle the temporal dimension of the input. Furthermore, the spiking neurons in SNNs have internal state variables, such as membrane potential, which necessitate storing data for each time step to update the membrane potential and execute the LIF process. Consequently, SNNs typically operate with smaller batch sizes compared to ANNs, which can lead to reduced pipeline performance in PipeLayer.

\begin{figure}[t]
    \centering
    \includegraphics[width=0.7\linewidth]{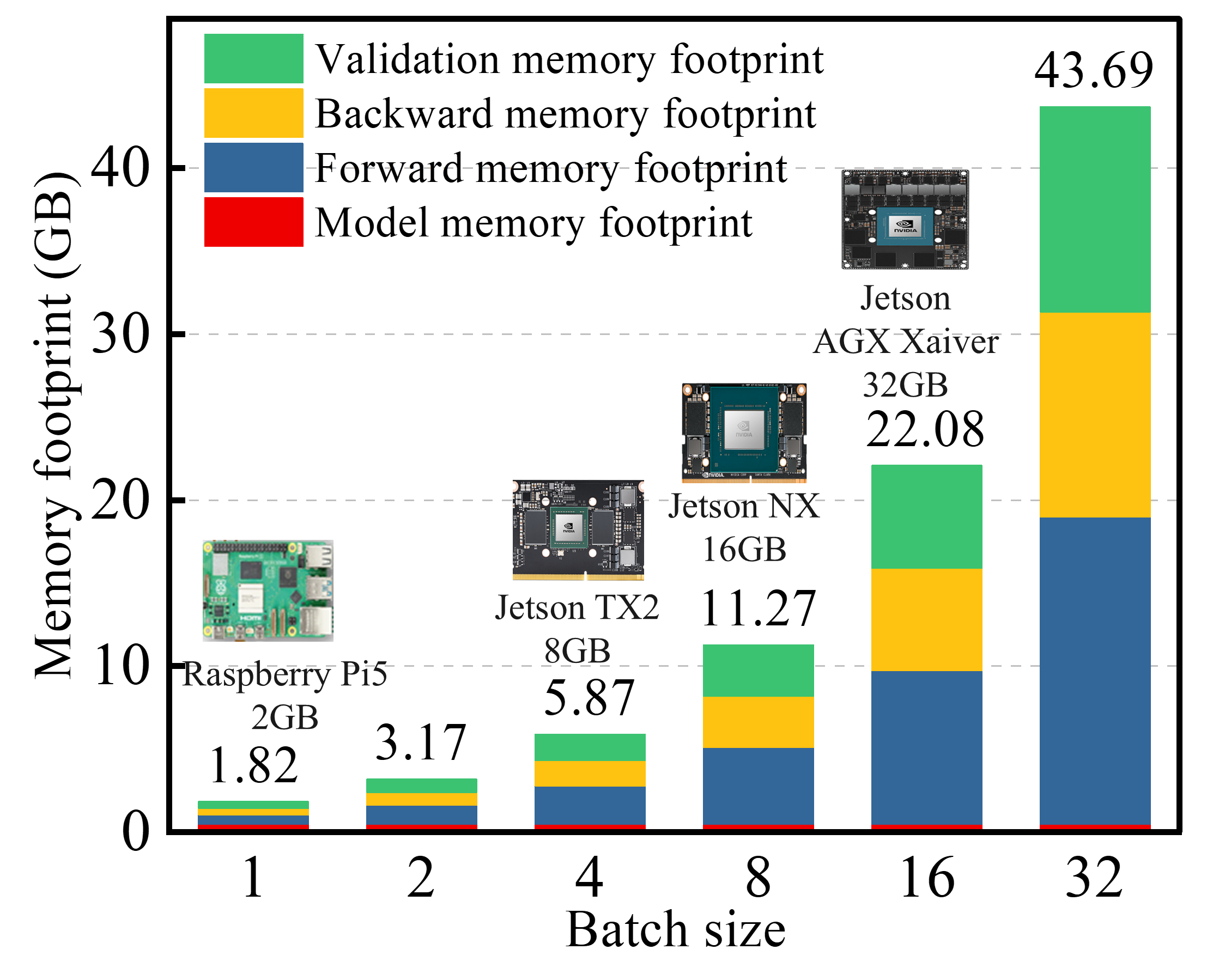}\
    \vspace{-10pt}
    \caption{Memory footprint of SNN training across batch sizes: requirements and corresponding edge devices at different capacity levels. PipeLayer suffers a significant decrease in acceleration efficiency under small batch size.}
    \label{fig:memory_model}
    \vspace{-10pt}
\end{figure}

\textbf{Limitation of PipeLayer.}
PipeLayer exploits both intra‑layer and inter‑layer parallelism to boost throughput \cite{pipelayer}. However, SNNs introduce an additional temporal dimension and per‑neuron state variables (e.g., membrane potentials) that must be stored and updated at every time step, driving up computational and memory overhead. Consequently, SNNs typically operate with smaller batch sizes than conventional ANNs, which in turn degrades PipeLayer’s pipeline efficiency. To quantify this trade‑off, we profiled the GPU memory footprint of VGG‑style SNN training across batch sizes from 1 to 32 \cite{vgg}. As shown in Fig. \ref{fig:memory_model}, memory requirements increase from modest usage at batch size 8 to over 32 GB at batch size 32, exceeding the 32 GB limit of edge platforms like the Jetson AGX Xavier. When the batch size is reduced from 32 to 8, PipeLayer exhibits a dramatic performance degradation of 55.5\%.

\textbf{Overview.}
To effectively accelerate training under small batch size conditions, we propose a timestep-data-batch three-level pipeline. Our pipelined dataflow architecture enhances SNN training efficiency by leveraging the temporal properties of SNNs and the hardware-friendly features introduced by SDFA. 

%At the finest granularity, the timestep serves as the smallest unit of data within a batch. By implementing a timestep-level pipeline, we significantly increase the effective operating frequency of the network. The data-level pipeline where each data sample within a batch is treated as a basic unit is the most common pipelining strategy in architectural design. However, due to the backward locking problem, BP-based architectures cannot support pipeline at the batch level. Our architecture overcomes this limitation, enabling batch-level pipelining that allows the system to fully benefit from smaller batch, deeper network depth, deeper temporal length.

\begin{figure}[!t]
    \centering
    \includegraphics[width=\linewidth]{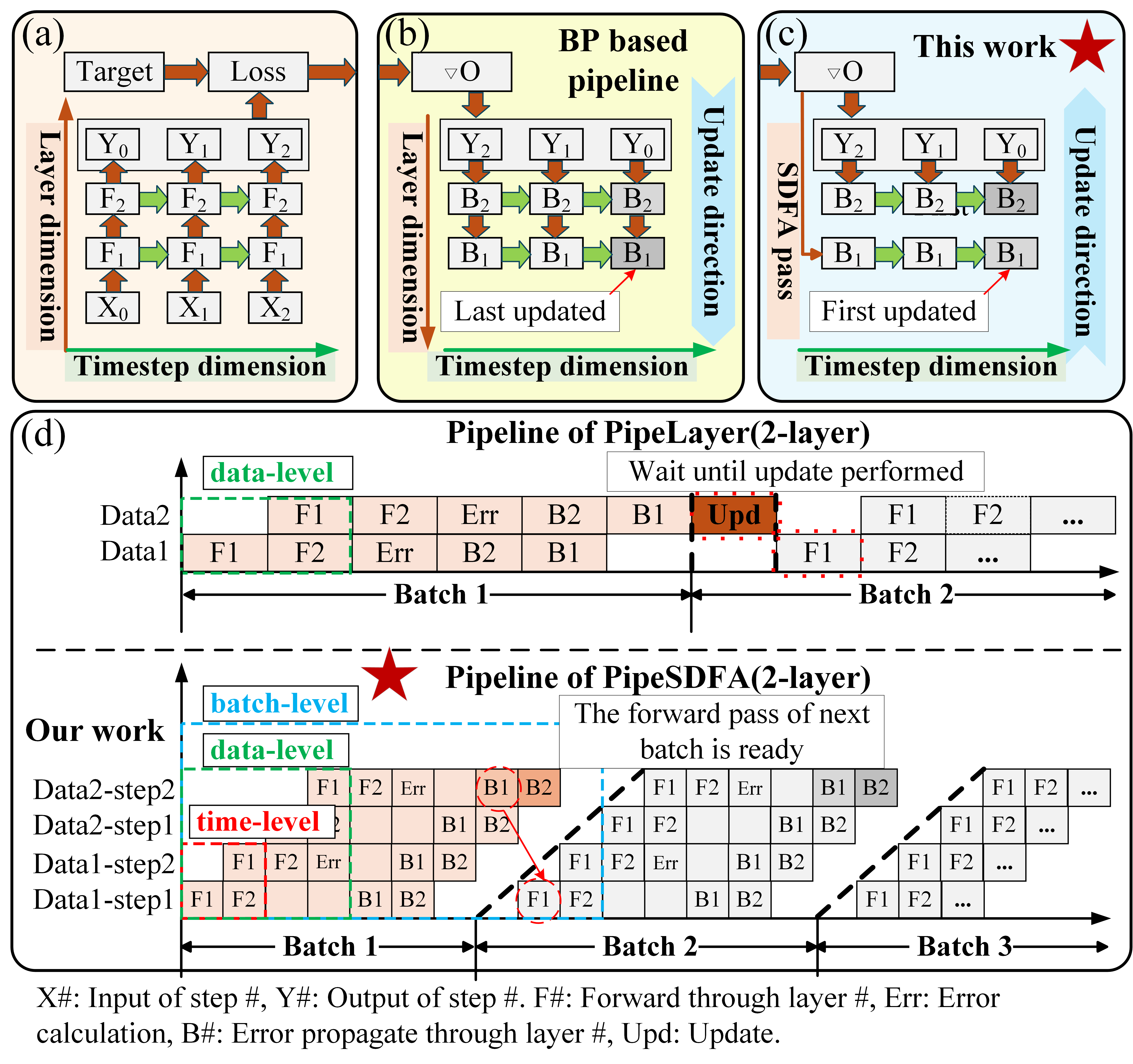}
    \vspace{-24pt}
    \caption{The timestep-level dataflow of (a) forward pass, (b) backward pass of BP, (c) backward pass of PipeSDFA with reversed update direction. (d) Data-level pipeline of PipeLayer vs. Three-level pipeline of PipeSDFA.}
    \label{fig:pipeline}
    \vspace{-21pt}
\end{figure}

\textbf{Timestep-level pipeline.}
Fig. \ref{fig:pipeline} (a-c) illustrates the timestep-level pipeline implementation for a two-layer SNN. Data are processed sequentially across both timesteps and layers, and the results are averaged at the output layer. To evaluate our pipeline's performance, we also implemented the timestep-level pipeline for BP, as depicted in Fig. \ref{fig:pipeline}(b). In BP, error gradients must be propagated along both the layer and timestep dimensions, and weight updates occur in the same direction as gradient propagation. Fig. \ref{fig:pipeline}(c) shows the pipeline design adapted to SDFA. As mentioned in Sec. \ref{sec:sdfa}, the SDFA algorithm enables direct propagation of the output layer error to intermediate layers, with the random feedback matrix $B\delta_L$ reused at each timestep. Therefore, the gradients for each layer in SDFA only need to be accumulated along timestep dimension, with the first layer weight gradients accumulating first. 

\textbf{Data-level pipeline.}
As discussed in Sec. \ref{sec:preliminaries}, input data are fed in batches during training, where each batch is processed using the same weights. Weight updates are first applied in the first layer after processing the entire batch. This property enables both pipelined and parallel processing within a batch. The green dashed box in Fig. \ref{fig:pipeline}(d) illustrates the implementation of data-level pipeline, assuming each data has two timesteps. The data-level pipeline implementation in our PipeSDFA is similar to that of PipeLayer \cite{pipelayer}. However, due to the finer-grained scheduling (time-step level pipeline), new input data are introduced into the pipeline every two cycles.

\textbf{Batch-level pipeline.}
The batch-level pipeline overlaps weight updates with the next batch's computations. The $Bi$ operation includes both error propagation and weight update calculations. Once the last timestep of the last data in a batch completes the $B1$ operation, and the next logical cycle can immediately start processing the next batch. Key to this efficiency is SDFA's property of timestep-invariant gradients (see line 17 in Alg. \ref{alg:sdfa}), which enables weight updates to commence immediately after processing the last timestep's $B1$ operation, rather than waiting for complete temporal processing as in BP. 

\textbf{PipeSDFA  vs. PipeLayer.} Table \ref{tab:cycle} compares the number of cycles required for both design approaches. For an input data, forward and backward pass require $(2L)$ cycles, with the weight update operation integrated within the error propagation process. Therefore, for a single data, the total number of required cycles is $(2L+T-1)$. For a batch of $B$ data, the total required cycles become $(2L+T+TB-1)$. Within a batch, the first weight update occurs at the $(L+T+TB)$-th cycle, after which ($N/B$) iterations are performed, and the total computation cycles can be expressed as $((L+T+TB)(N/B)+L-1)$.

\begin{table}[h]
    \centering
    \vspace{-10pt}
    \caption{Cycle Comparison: PipeLayer vs. Our PipeSDFA.}
    \vspace{-6pt}
    \begin{tabular}{|c|c|}
        \hline
        \multicolumn{2}{|c|}{L: Number of layers;   T: Timestep of input} \\ 
        \multicolumn{2}{|c|}{B: Batch size;   N: Total Number of input} \\ \hline
        \hline
        \textbf{Method}& \textbf{Cycle} \\
        \hline
        PipeLayer w/o pipeline \cite{pipelayer} & [(2L+1)N + N/B]T \\ \hline
        PipeLayer with pipeline \cite{pipelayer}& (N/B)(2L+B+1)T \\ \hline
        Our PipeSDFA & (L+T+TB)(N/B)+L-1 \\
        \hline
    \end{tabular}
    \vspace{-10pt}
    \label{tab:cycle}
\end{table}

\begin{table*}[ht]
    \centering
    \caption{Model setup and accuracy evaluation of SDFA.}
    \vspace{-4pt}
    \resizebox{0.95\textwidth}{!}{%
        \begin{tabular}{ccccccc}
            \toprule % Use booktabs rules if available
            \multirow{2}{*}{\textbf{Dataset}} & \multirow{2}{*}{\textbf{Number of classes}} & \multirow{2}{*}{\textbf{Neuron type}} & \multirow{2}{*}{\textbf{Model}} & \multirow{2}{*}{\textbf{Setup}} &\multicolumn{2}{c}{\textbf{Accuracy(\%)}} \\
            \cmidrule(lr){6-7} % Use cmidrule for partial rule under multicolumn
            & & & & & \textbf{BP} & \textbf{SDFA (This work)} \\
            \midrule % Use booktabs rules if available
            %--- N-MNIST Group (Gray) ---
            \rowcolor{lightgray} % Apply color to row 1
             &  &  & MLP-A & 2 $\times$ FC & 98.52 & \textbf{97.92} \\
            \rowcolor{lightgray} % Apply color to row 2
             &  &  & MLP-B & 3 $\times$ FC & 98.46 & \textbf{97.97} \\
            \rowcolor{lightgray} % Apply color to row 3
             &  &  & MLP-C & 4 $\times$ FC & 98.38 & \textbf{97.95} \\
            \rowcolor{lightgray} % Apply color to row 4
            % *** FIX: Place multirow on LAST row with negative span ***
            \multirow{-4}{*}{N-MNIST \cite{nmnist}} & \multirow{-4}{*}{10} & \multirow{-4}{*}{LIF} & MLP-D & 5 $\times$ FC & 97.87 & \textbf{97.82} \\
            %\midrule % Use booktabs rules if available
            %--- SHD Group (White) ---
            % No \rowcolor needed for white
            SHD \cite{shd}& 20 & IF & MLP-B & 3 $\times$ FC & 74.09 & \textbf{73.29} \\
            %\midrule % Use booktabs rules if available
            %--- Braille Group (Gray) ---
            \rowcolor{lightgray}
            Braille letter \cite{braille} & 28 & IF & MLP-B & 3 $\times$ FC & 99.62 & \textbf{99.26} \\
            %\midrule % Use booktabs rules if available
            %--- DVS-Gesture Group (White) ---
            % No \rowcolor needed for white (row 1)
             &  &  & ConvNet &  5 $\times$ Conv + 2 $\times$ FC & 95.14 & \textbf{93.75} \\
             % No \rowcolor needed for white (row 2)
             % *** FIX: Place multirow on LAST row with negative span ***
            \multirow{-2}{*}{DVS-Gesture \cite{dvs}} & \multirow{-2}{*}{11} & \multirow{-2}{*}{IF} & VGG11 & 8 $\times$ Conv + 3 $\times$ FC & 88.92 & \textbf{92.36} \\
            %\midrule % Use booktabs rules if available
            %--- N-Caltech101 Group (Gray) ---
            \rowcolor{lightgray}
            N-Caltech101 \cite{nmnist} & 101 & IF & VGG11 & 8 $\times$ Conv + 3 $\times$ FC & 59.53 & \textbf{64.38} \\
            \bottomrule % Use booktabs rules if available
            \multicolumn{7}{l}{FC: Fully Connected Layer; Conv: Convolutional Layer; LIF: Leaky Integrate-and-Fire; IF: Integrate-and-Fire.} \\ % Use \multicolumn{7}{l} for left alignment
        \end{tabular}%
    }
    \label{tab:model_acc}
    \vspace{-6pt}
\end{table*}

\section{EXPERIMENTAL RESULTS}

\subsection{Experiment Setup}

\textbf{Software Benchmarks.} We conducted experiments using four Multi-Layer Perceptrons (MLPs) and two Convolutional Neural Networks (CNNs) as shown in Table \ref{tab:model_acc} on the N-MNIST \cite{nmnist}, DVS-Gesture \cite{dvs}, Spiking Heidelberg Dataset \cite{shd}, Braille letter \cite{braille} and N-Caltech101 \cite{nmnist} to evaluate our proposed SDFA algorithm and conducted five repeated runs for statistical robustness. To compare effectiveness of the SDFA, we used BP-based method to train models on various datasets as our baselines. 
%In practical scenarios, RRAM cells can only support limited precision levels. We conducted a series of experiments to investigate the trade-off between RRAM cell resolution and accuracy.

% To evaluate the impact of quantization error on the edge devices, we used the floating-point weights of the network trained by SDFA as the baseline and assess the trade-off between resolution and accuracy.

\textbf{Hardware Accelerators.} We build a cycle-accurate simulator to assess the effectiveness of the in-memory training architecture. For fair comparison, we adopted the same settings as PipeLayer, with the RRAM read/write latency set to 29.31 ns/50.88 ns, and the energy consumption per spike set to 1.08 pJ/3.91 nJ respectively \cite{pipelayer}. The RRAM array size is fixed at 256 $\times$ 256. We developed a Python based toolchain to evaluate the speedup and energy saving ratio with the BP-based PipeLayer architecture \cite{pipelayer} and ours. We evaluated both architectures using the same mapping strategy (VW-SDK \cite{vwsdk}). %Evaluations were performed on both Spiking MLPs and CNNs.

\subsection{Evaluations of SDFA}
\textbf{Accuracy.}
The proposed SDFA method achieves consistently high accuracy across almost all neuromorphic datasets, as shown in Table \ref{tab:model_acc}. On standard benchmarks such as N-MNIST, SHD, and Braille letter, SDFA maintains accuracy comparable to the BP algorithm, with an accuracy gap of no more than 2\% across various network depths and architectures. This demonstrates that SDFA is robust and effective, even as network complexity increases from simple MLPs to deeper architectures.
Specifically, for MLPs with two to five layers on the N-MNIST dataset, SDFA achieves nearly identical performance to BP, indicating that its learning dynamics are effective for both shallow and deep configurations. Similarly, on the SHD and Braille letter datasets, which represent more specialized recognition tasks, SDFA delivers high accuracy, further highlighting its adaptability to different data characteristics.

A particularly noteworthy phenomenon is observed on more complex datasets and architectures. On the DVS-Gesture and N-Caltech101 datasets, SDFA not only maintains competitive performance but even surpasses BP when evaluated using VGG11 architecture. For example, on N-Caltech101, SDFA achieves 64.38\% accuracy compared to 59.53\% for BP, and on DVS-Gesture, SDFA reaches 92.36\% versus 88.92\% for BP. This superior performance may be attributed to the unique learning dynamics of SDFA, which could provide regularization benefits and enhanced generalization in complex scenarios.

\begin{figure}[t]
    \centering
    \includegraphics[width=\linewidth]{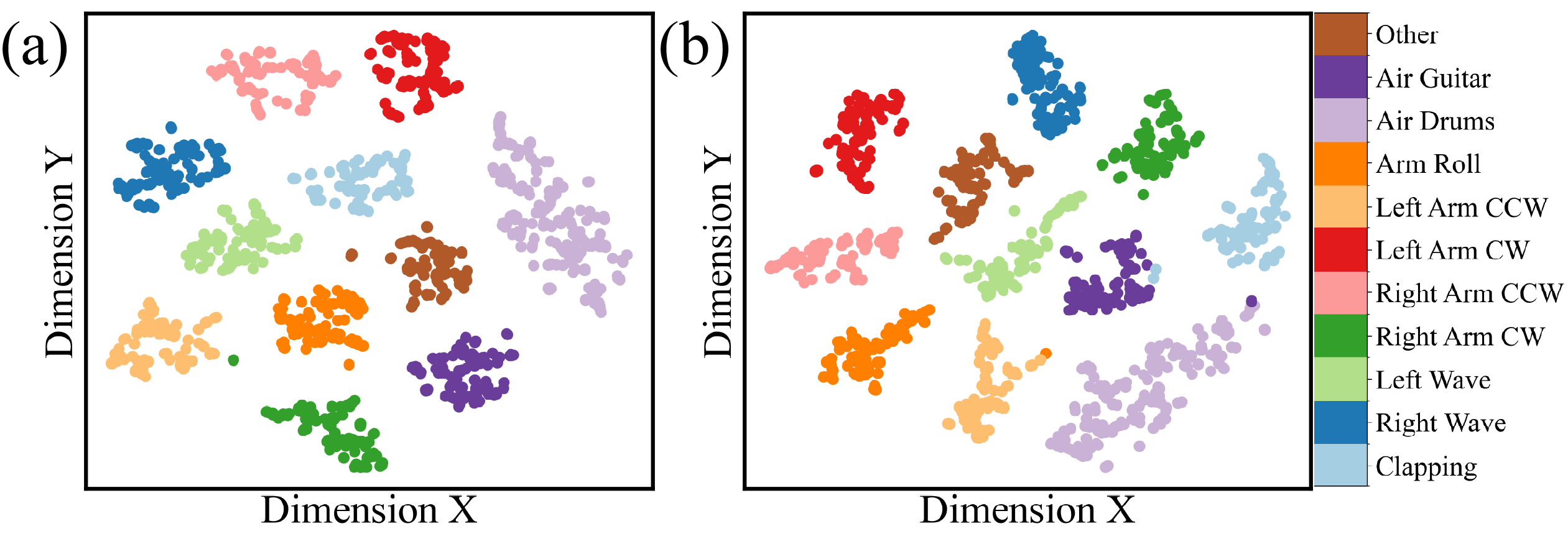}
    \vspace{-23pt}
    \caption{T-SNE visualization of latent features for DVS-Gesture dataset: (a) BP vs. (b) SDFA.}
    \label{fig:tsne}\
    \vspace{-18pt}
\end{figure}

\textbf{T-SNE analysis.}
The t-SNE analysis in Fig. \ref{fig:tsne} provides compelling visual evidence of the feature learning capabilities of SDFA compared to BP on the DVS-Gesture dataset. Both methods demonstrate effective class separation, with distinct clusters forming for different gesture categories. This visualization confirms that despite its simplified feedback mechanism, SDFA learns discriminative spatiotemporal representations that capture the essential dynamics of event-based gesture data, aligning with the quantitative accuracy results reported in Table \ref{tab:model_acc}. The comparable clustering performance between both methods further validates SDFA as a viable alternative to BP for SNN training.

\begin{figure}[t]
    \centering
    \includegraphics[width=\linewidth]{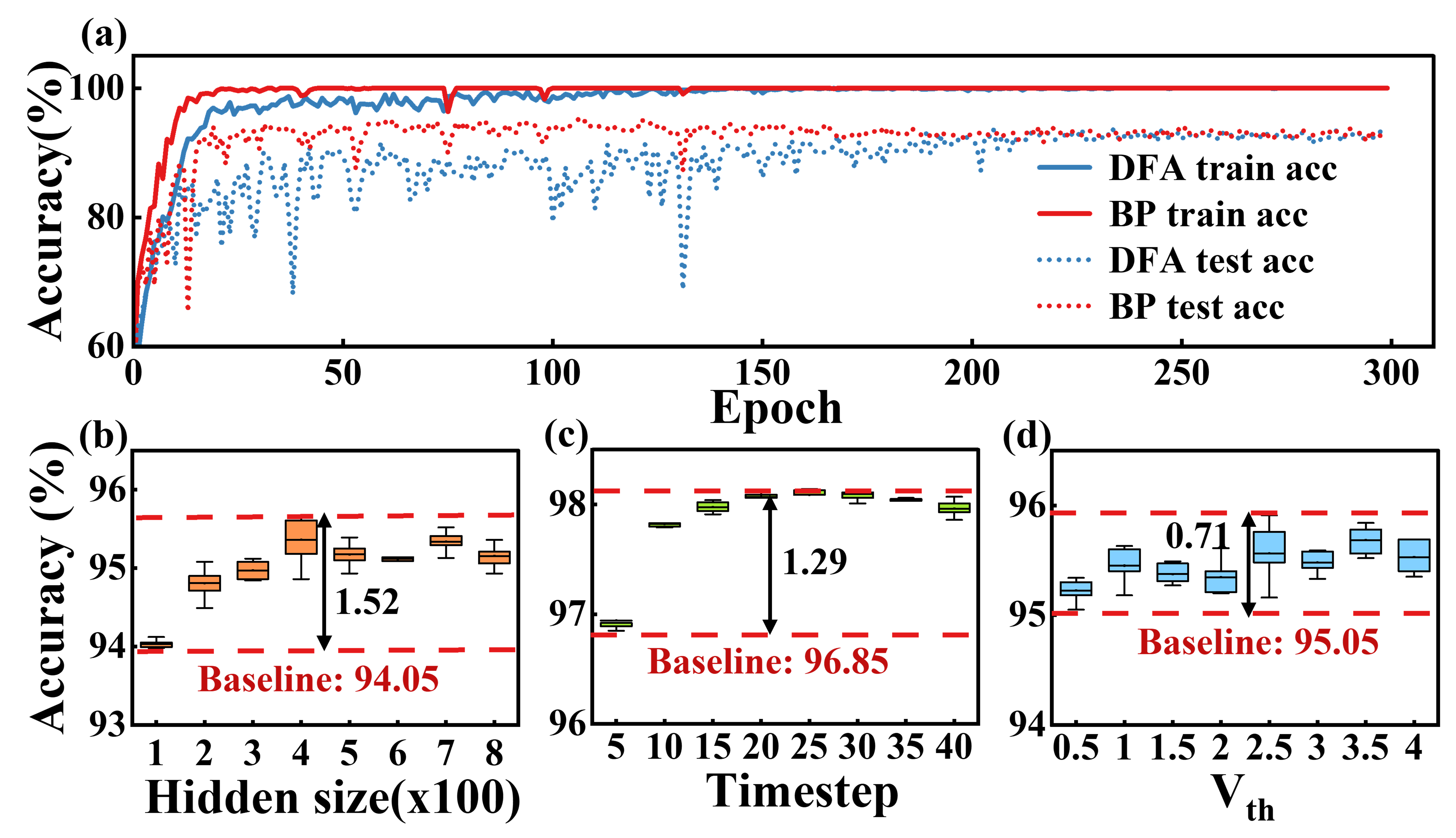}
    \vspace{-16pt}
    \caption{(a) Convergence speed of SDFA vs. BP. Hyperparameter exploration, (b) hidden size, (c) timestep, (d) $V_{th}$.}
    \label{fig:convergency}
    \vspace{-16pt}
\end{figure}

\textbf{Convergence speed.}
We evaluated the convergence speed of SDFA on the DVS-Gesture dataset using ConvNet, with results shown in Fig. \ref{fig:convergency}(a). The convergence plots demonstrate that SDFA exhibits comparable learning dynamics to BP. Though SDFA converges slightly more slowly than BP and exhibits more pronounced fluctuations on the test set, it maintains a steady improvement trajectory throughout training. Around 150 epochs, SDFA achieves convergence with an accuracy comparable to BP. This slightly slower convergence speed is effectively offset by the significant hardware acceleration benefits that SDFA enables. The elimination of the backward locking problem allows for highly parallelized error propagation and weight updates, resulting in per-iteration speedups that more than compensate for the additional epochs required. This favorable trade-off between algorithmic convergence and hardware efficiency underscores SDFA's practical value for edge computing applications.

\textbf{Hyperparameter exploration.}
We explored the impact of key hyperparameters, including hidden size, timestep, and threshold voltage ($V_{th}$). As shown in Fig. \ref{fig:convergency}(b-d), SDFA demonstrates robust performance across a wide range of configurations, underscoring the algorithm’s flexibility and broad design applicability.
Specifically, increasing the hidden size leads to improved accuracy. As the number of timesteps increases, both network stability and accuracy are enhanced. However, beyond 25 timesteps, further increases produce a slight negative effect on accuracy, indicating that excessive time steps are not conducive to classification tasks. Notably, SDFA maintains stable accuracy over a broad range of $V_{th}$.

\subsection{Evaluation of PipeSDFA}

\begin{figure*}[t]
    \centering
    \includegraphics[width=\linewidth]{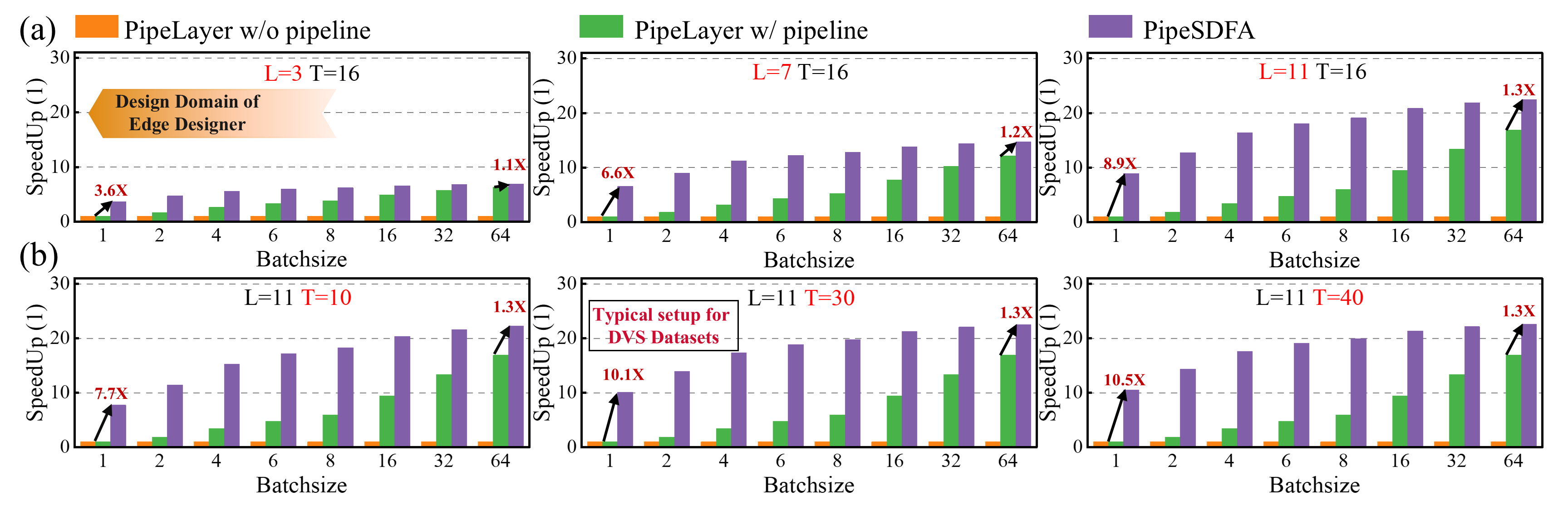}
    \vspace{-20pt}
    \caption{Batch size impact on speedup ratio of PipeLayer vs. Pipelined SDFA across (a) layers, and (b) timesteps.}
    \label{fig:speedup}
    \vspace{-14pt}
\end{figure*}

\textbf{Speedup.}
Our comprehensive performance evaluation demonstrates that PipeSDFA achieves substantial speedup compared to PipeLayer architecture across various network configurations and operational parameters. Fig. \ref{fig:speedup} illustrates the speedup factors of 1.1$\times$$\sim$10.5$\times$ over PipeLayer, with performance gains highly dependent on network topology and training parameters.

As shown in Fig. \ref{fig:speedup} (a)-(b), PipeSDFA exhibits remarkable scaling efficiency as both network depth (L) and time steps (T) increase. For instance, at L=7 and T=16, we observe a speedup of 6.6$\times$ over PipeLayer with 1 batch size, which further increases to 8.9$\times$ when L increases to 10. This performance improvement with increasing network complexity highlights PipeSDFA's effectiveness for modern SNN applications requiring deeper architectures.

Edge computing applications typically operate with constrained batch size due to memory limitations and real-time processing requirements. Our analysis reveals that PipeSDFA maintains significant performance advantages in this critical design space. For typical setups using the DVS-Gesture dataset, where batch sizes rarely exceed 8 due to the event-driven nature of the data, PipeSDFA consistently achieves a minimum speedup of 4.3$\times$ over PipeLayer. This performance advantage is particularly valuable for real-time visual processing applications on edge devices, where both throughput and energy efficiency are critical constraints.

\begin{figure}[t]
    \centering
    \includegraphics[width=0.95\linewidth]{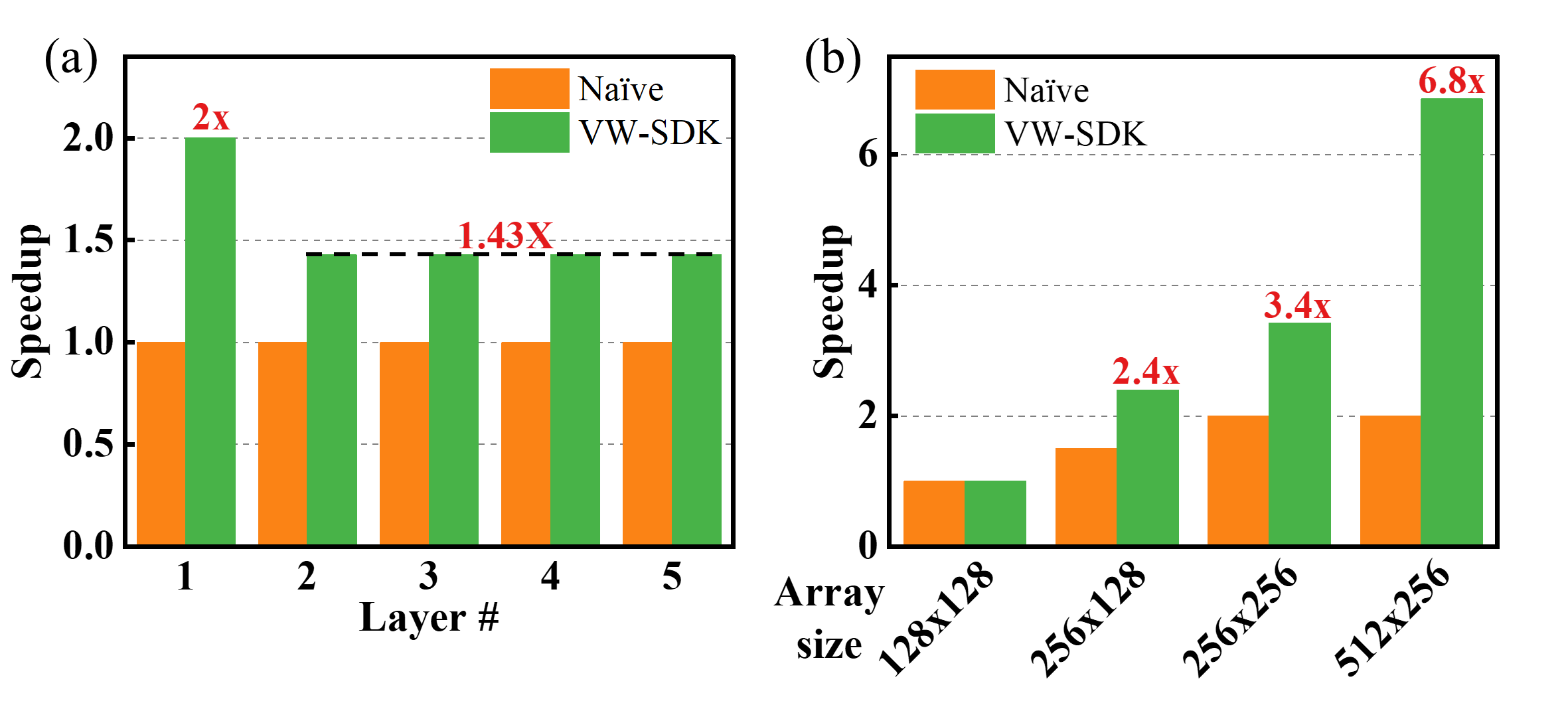}
    \vspace{-16pt}
    \caption{Comparison of the normalized speedup (a) for convolutional layer of ConvNet using $256 \times 256$ array, (b) for the entire layers of ConvNet.}
    \label{fig:sdk}
    \vspace{-16pt}
\end{figure}

\textbf{Mapping strategy.}
Fig. \ref{fig:sdk} demonstrates the acceleration benefits of the VW-SDK mapping strategy compared to direct mapping without input data reuse. The results show that this approach takes full advantage of large arrays and significantly improves both the utilization and processing speed of convolutional layers in ConvNet.

\begin{figure}[t]
    \centering
    \includegraphics[width=0.95\linewidth]{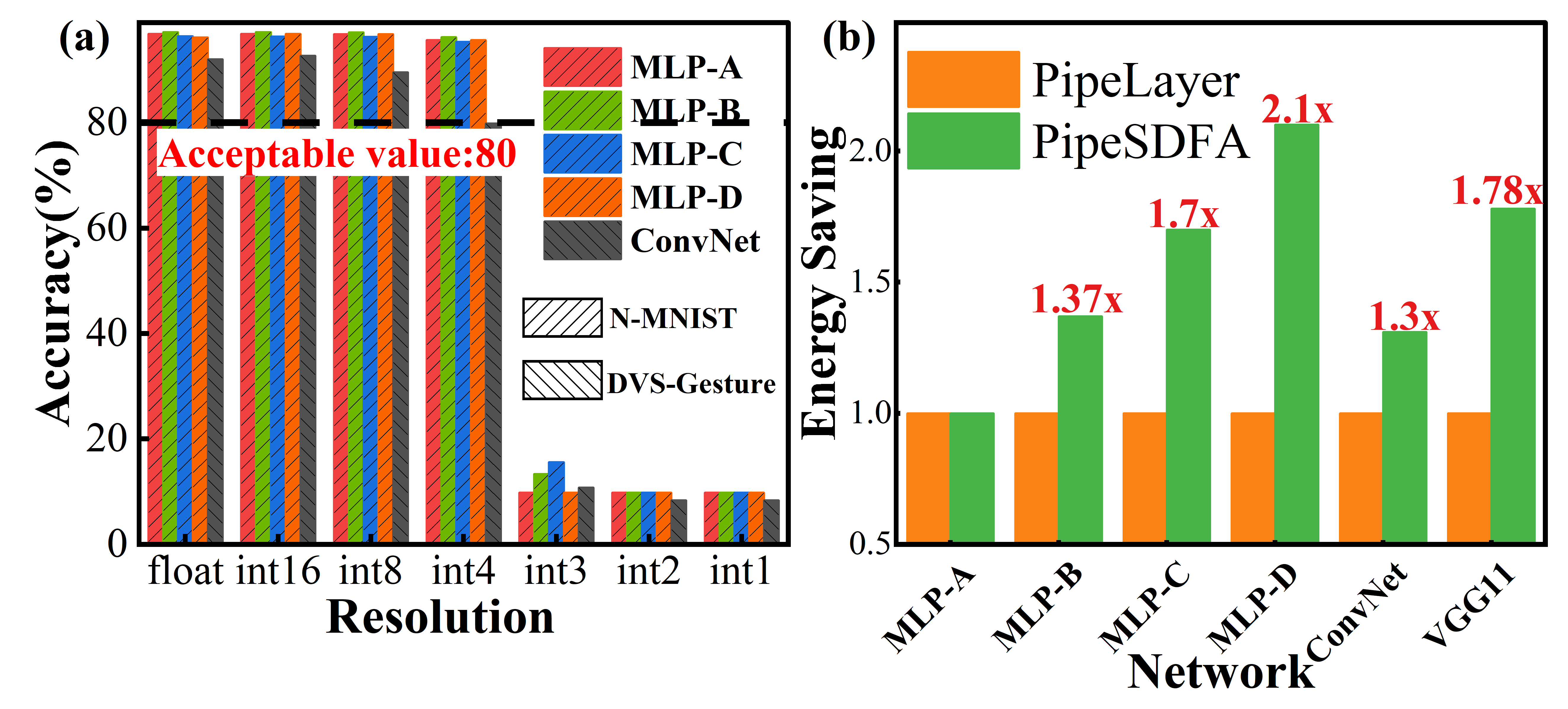}
    \vspace{-6pt}
    \caption{(a) Quantization evaluation. (b) Energy saving: PipeLayer vs. SDFA.}
    \label{fig:quant_energy}
    \vspace{-16pt}
\end{figure}

\textbf{Quantization.}
To determine the optimal bit precision for RRAM cells in our architecture, we conducted extensive quantization experiments across five network structures. As illustrated in Fig. \ref{fig:quant_energy}(a), accuracy remains stable when network precision exceeds 4 bits. At the critical 4-bit threshold, even the most sensitive model (ConvNet) maintains approximately 80\%. This finding is particularly significant for RRAM-based implementations, as it enables the use of compact 4-bit cells without substantial performance penalties, reducing both area and energy requirements while maintaining competitive accuracy on neuromorphic datasets.

\textbf{Energy.}
Fig. \ref{fig:quant_energy}(b) quantifies the energy advantages of PipeSDFA over PipeLayer. PipeSDFA achieves a substantial 1.37$\times$$\sim$2.1$\times$ reduction in energy consumption, primarily attributable to our optimized error propagation mechanism and three-level pipeline design. While both architectures exhibit comparable energy profiles during forward propagation and weight update phases, our streamlined backward pass and reduced buffer requirements translate to substantial energy savings at the chip level. This efficiency gain is particularly valuable for edge applications where energy constraints often determine deployment feasibility.

\section{Conclusion}

We propose a novel software-hardware co-design for rapid and energy-efficient SNN training of event data on the edge. Software-wise, we introduce a hardware-friendly training algorithm SDFA, which effectively reduces the computational complexity of SNN training by eliminating sequential error propagation. Hardware-wise, we design a three-level pipelined dataflow based on IMC architecture PipeSDFA, which parallelize the training process. Evaluations demonstrate that the PipeSDFA training accelerator surpasses traditional BP-based accelerators in both performance and efficiency, providing a scalable solution for the application of neuromorphic systems.
\section{ACKNOWLEDGEMENT}
This research is supported by the Strategic Priority Research Program of the Chinese Academy of Sciences under Grant No. XDA0330100, the National Natural Science Foundation of China (Grant Nos. 62374181, 62322412, 92464201, U2341218), Joint Laboratory of Microelectronics (JLFS/E-601/24), Beijing Natural Science Foundation (Grant No. Z210006), Hong Kong Research Grant Council (Grant No. 17212923), Shenzhen Science and Technology Innovation Commission (SGDX20220530111405040).

\clearpage
\small
\bibliography{references}

\begin{thebibliography}{10}
\providecommand{\url}[1]{#1}
\csname url@samestyle\endcsname
\providecommand{\newblock}{\relax}
\providecommand{\bibinfo}[2]{#2}
\providecommand{\BIBentrySTDinterwordspacing}{\spaceskip=0pt\relax}
\providecommand{\BIBentryALTinterwordstretchfactor}{4}
\providecommand{\BIBentryALTinterwordspacing}{\spaceskip=\fontdimen2\font plus
\BIBentryALTinterwordstretchfactor\fontdimen3\font minus \fontdimen4\font\relax}
\providecommand{\BIBforeignlanguage}[2]{{%
\expandafter\ifx\csname l@#1\endcsname\relax
\typeout{** WARNING: IEEEtran.bst: No hyphenation pattern has been}%
\typeout{** loaded for the language `#1'. Using the pattern for}%
\typeout{** the default language instead.}%
\else
\language=\csname l@#1\endcsname
\fi
#2}}
\providecommand{\BIBdecl}{\relax}
\BIBdecl

\bibitem{peiArtificial2019}
J.~Pei, L.~Deng, S.~Song, M.~Zhao, Y.~Zhang, S.~Wu \emph{et~al.}, ``Towards artificial general intelligence with hybrid {{Tianjic}} chip architecture,'' \emph{Nature}, vol. 572, no. 7767, pp. 106--111, Aug. 2019.

\bibitem{tianNew2021}
F.~Tian, J.~Yang, S.~Zhao, and M.~Sawan, ``A {{New Neuromorphic Computing Approach}} for {{Epileptic Seizure Prediction}},'' in \emph{2021 {{IEEE International Symposium}} on {{Circuits}} and {{Systems}} ({{ISCAS}})}, May 2021, pp. 1--5.

\bibitem{doborjehPersonalised2021}
M.~Doborjeh, Z.~Doborjeh, A.~Merkin, H.~Bahrami, A.~Sumich, R.~Krishnamurthi \emph{et~al.}, ``Personalised predictive modelling with brain-inspired spiking neural networks of longitudinal {{MRI}} neuroimaging data and the case study of dementia,'' \emph{Neural Networks}, vol. 144, pp. 522--539, Dec. 2021.

\bibitem{LNPU}
D.~Han, J.~Lee, and H.-J. Yoo, ``{{DF-LNPU}}: {{A Pipelined Direct Feedback Alignment-Based Deep Neural Network Learning Processor}} for {{Fast Online Learning}},'' \emph{IEEE Journal of Solid-State Circuits}, vol.~56, no.~5, pp. 1630--1640, May 2021.

\bibitem{bp}
D.~E. Rumelhart, G.~E. Hinton, and R.~J. Williams, ``Learning internal representations by error propagation, parallel distributed processing, explorations in the microstructure of cognition, ed. de rumelhart and j. mcclelland. vol. 1. 1986,'' \emph{Biometrika}, vol.~71, no. 599-607, p.~6, 1986.

\bibitem{ISAAC}
A.~Shafiee, A.~Nag, N.~Muralimanohar, R.~Balasubramonian, J.~P. Strachan, M.~Hu \emph{et~al.}, ``{{ISAAC}}: {{A Convolutional Neural Network Accelerator}} with {{In-Situ Analog Arithmetic}} in {{Crossbars}},'' in \emph{2016 {{ACM}}/{{IEEE}} 43rd {{Annual International Symposium}} on {{Computer Architecture}} ({{ISCA}})}, Jun. 2016, pp. 14--26.

\bibitem{linResistive2025}
N.~Lin, S.~Wang, Y.~Li, B.~Wang, S.~Shi, Y.~He \emph{et~al.}, ``Resistive memory-based zero-shot liquid state machine for multimodal event data learning,'' \emph{Nature Computational Science}, vol.~5, no.~1, pp. 37--47, Jan. 2025.

\bibitem{inference}
X.~Peng, R.~Liu, and S.~Yu, ``Optimizing {{Weight Mapping}} and {{Data Flow}} for {{Convolutional Neural Networks}} on {{Processing-in-Memory Architectures}},'' \emph{IEEE Transactions on Circuits and Systems I: Regular Papers}, vol.~67, no.~4, pp. 1333--1343, Apr. 2020.

\bibitem{pipelayer}
L.~Song, X.~Qian, H.~Li, and Y.~Chen, ``{{PipeLayer}}: {{A Pipelined ReRAM-Based Accelerator}} for {{Deep Learning}},'' in \emph{2017 {{IEEE International Symposium}} on {{High Performance Computer Architecture}} ({{HPCA}})}.\hskip 1em plus 0.5em minus 0.4em\relax Austin, TX: IEEE, Feb. 2017, pp. 541--552.

\bibitem{yuRRAM2021}
S.~Yu, W.~Shim, X.~Peng, and Y.~Luo, ``{{RRAM}} for {{Compute-in-Memory}}: {{From Inference}} to {{Training}},'' \emph{IEEE Transactions on Circuits and Systems I: Regular Papers}, vol.~68, no.~7, pp. 2753--2765, Jul. 2021.

\bibitem{locking}
M.~Jaderberg, W.~M. Czarnecki, S.~Osindero, O.~Vinyals, A.~Graves, D.~Silver \emph{et~al.}, ``Decoupled neural interfaces using synthetic gradients,'' in \emph{International Conference on Machine Learning}.\hskip 1em plus 0.5em minus 0.4em\relax PMLR, 2017, pp. 1627--1635.

\bibitem{hanExtension2020}
D.~Han, G.~Park, J.~Ryu, and H.-j. Yoo, ``Extension of {{Direct Feedback Alignment}} to {{Convolutional}} and {{Recurrent Neural Network}} for {{Bio-plausible Deep Learning}},'' Jun. 2020.

\bibitem{transpose1}
J.-W. Su, X.~Si, Y.-C. Chou, T.-W. Chang, W.-H. Huang, Y.-N. Tu \emph{et~al.}, ``15.2 {{A}} 28nm {{64Kb Inference-Training Two-Way Transpose Multibit 6T SRAM Compute-in-Memory Macro}} for {{AI Edge Chips}},'' in \emph{2020 {{IEEE International Solid- State Circuits Conference}} - ({{ISSCC}})}.\hskip 1em plus 0.5em minus 0.4em\relax San Francisco, CA, USA: IEEE, Feb. 2020, pp. 240--242.

\bibitem{transpose2}
C.~Kim, S.~Kang, D.~Shin, S.~Choi, Y.~Kim, and H.-J. Yoo, ``A 2.{{1TFLOPS}}/{{W Mobile Deep RL Accelerator}} with {{Transposable PE Array}} and {{Experience Compression}},'' in \emph{2019 {{IEEE International Solid-State Circuits Conference}} - ({{ISSCC}})}, Feb. 2019, pp. 136--138.

\bibitem{transpose3}
J.~Lee, S.~Kim, S.~Kim, W.~Jo, D.~Han, J.~Lee \emph{et~al.}, ``{{OmniDRL}}: {{A}} 29.3 {{TFLOPS}}/{{W Deep Reinforcement Learning Processor}} with {{Dualmode Weight Compression}} and {{On-chip Sparse Weight Transposer}},'' in \emph{2021 {{Symposium}} on {{VLSI Circuits}}}, Jun. 2021, pp. 1--2.

\bibitem{dfa}
A.~N{\o}kland, ``Direct feedback alignment provides learning in deep neural networks,'' \emph{Advances in neural information processing systems}, vol.~29, 2016.

\bibitem{dfa_deeper}
J.~Launay, I.~Poli, F.~Boniface, and F.~Krzakala, ``Direct {Feedback} {Alignment} {Scales} to {Modern} {Deep} {Learning} {Tasks} and {Architectures},'' in \emph{Advances in {Neural} {Information} {Processing} {Systems}}, vol.~33.\hskip 1em plus 0.5em minus 0.4em\relax Curran Associates, Inc., 2020, pp. 9346--9360.

\bibitem{snn_dfa1}
S.~Bang, D.~Lew, S.~Choi, and J.~Park, ``An energy-efficient {{SNN}} processor design based on sparse direct feedback and spike prediction,'' in \emph{2021 {{International Joint Conference}} on {{Neural Networks}} ({{IJCNN}})}.\hskip 1em plus 0.5em minus 0.4em\relax IEEE, 2021, pp. 1--8.

\bibitem{snn_dfa2}
C.~Shi, T.~Wang, J.~He, J.~Zhang, L.~Liu, and N.~Wu, ``Deeptempo: A hardware-friendly direct feedback alignment multi-layer tempotron learning rule for deep spiking neural networks,'' \emph{IEEE Transactions on Circuits and Systems II: Express Briefs}, vol.~68, no.~5, pp. 1581--1585, 2021.

\bibitem{snn_dfa3}
J.~Lee, R.~Zhang, W.~Zhang, Y.~Liu, and P.~Li, ``Spike-{{Train Level Direct Feedback Alignment}}: {{Sidestepping Backpropagation}} for {{On-Chip Training}} of {{Spiking Neural Nets}},'' \emph{Frontiers in Neuroscience}, vol.~14, p. 143, Mar. 2020.

\bibitem{snn_dfa4}
Y.~Zhang, K.~Inoue, M.~Nakajima, T.~Hashimoto, Y.~Kuniyoshi, and K.~Nakajima, ``Training {{Spiking Neural Networks}} via {{Augmented Direct Feedback Alignment}},'' Sep. 2024.

\bibitem{lif1}
M.~Doborjeh, Z.~Doborjeh, A.~Merkin, H.~Bahrami, A.~Sumich, R.~Krishnamurthi \emph{et~al.}, ``Personalised predictive modelling with brain-inspired spiking neural networks of longitudinal {{MRI}} neuroimaging data and the case study of dementia,'' \emph{Neural Networks}, vol. 144, pp. 522--539, Dec. 2021.

\bibitem{stdp1}
G.-q. Bi and M.-m. Poo, ``Synaptic modifications in cultured hippocampal neurons: Dependence on spike timing, synaptic strength, and postsynaptic cell type,'' \emph{Journal of neuroscience}, vol.~18, no.~24, pp. 10\,464--10\,472, 1998.

\bibitem{stdp2}
P.~U. Diehl and M.~Cook, ``Unsupervised learning of digit recognition using spike-timing-dependent plasticity,'' \emph{Frontiers in computational neuroscience}, vol.~9, p.~99, 2015.

\bibitem{ann2snn}
B.~Rueckauer, I.-A. Lungu, Y.~Hu, M.~Pfeiffer, and S.-C. Liu, ``Conversion of continuous-valued deep networks to efficient event-driven networks for image classification,'' \emph{Frontiers in neuroscience}, vol.~11, p. 682, 2017.

\bibitem{surrogate2}
S.~B. Shrestha and G.~Orchard, ``Slayer: {{Spike}} layer error reassignment in time,'' \emph{Advances in neural information processing systems}, vol.~31, 2018.

\bibitem{surrogate3}
F.~Zenke and S.~Ganguli, ``Superspike: {{Supervised}} learning in multilayer spiking neural networks,'' \emph{Neural computation}, vol.~30, no.~6, pp. 1514--1541, 2018.

\bibitem{vwsdk}
J.~Rhe, S.~Moon, and J.~H. Ko, ``{{VW-SDK}}: {{Efficient Convolutional Weight Mapping Using Variable Windows}} for {{Processing-In-Memory Architectures}},'' in \emph{2022 {{Design}}, {{Automation}} \& {{Test}} in {{Europe Conference}} \& {{Exhibition}} ({{DATE}})}.\hskip 1em plus 0.5em minus 0.4em\relax Antwerp, Belgium: IEEE, Mar. 2022, pp. 214--219.

\bibitem{vgg}
K.~Simonyan, ``Very deep convolutional networks for large-scale image recognition,'' \emph{arXiv preprint arXiv:1409.1556}, 2014.

\bibitem{nmnist}
G.~Orchard, A.~Jayawant, G.~K. Cohen, and N.~Thakor, ``Converting static image datasets to spiking neuromorphic datasets using saccades,'' \emph{Frontiers in neuroscience}, vol.~9, p. 437, 2015.

\bibitem{shd}
B.~Cramer, Y.~Stradmann, J.~Schemmel, and F.~Zenke, ``The {{Heidelberg}} spiking datasets for the systematic evaluation of spiking neural networks,'' \emph{IEEE Transactions on Neural Networks and Learning Systems}, vol.~33, no.~7, pp. 2744--2757, Jul. 2022.

\bibitem{braille}
S.~F. {Muller-Cleve}, V.~Fra, L.~Khacef, A.~{Pequeno-Zurro}, D.~Klepatsch, E.~Forno \emph{et~al.}, ``Braille {{Letter Reading}}: {{A Benchmark}} for {{Spatio-Temporal Pattern Recognition}} on {{Neuromorphic Hardware}},'' Oct. 2022.

\bibitem{dvs}
A.~Amir, B.~Taba, D.~Berg, T.~Melano, J.~McKinstry, C.~Di~Nolfo \emph{et~al.}, ``A low power, fully event-based gesture recognition system,'' in \emph{Proceedings of the {{IEEE}} Conference on Computer Vision and Pattern Recognition}, 2017, pp. 7243--7252.

\end{thebibliography}

\end{document}